\documentclass[review]{elsarticle}

\usepackage{lineno,hyperref}
\modulolinenumbers[5]
\usepackage{amsmath}
\usepackage{braket}
\usepackage{algorithm}%
\usepackage{algorithmicx}%
\usepackage{algpseudocode}%

\usepackage[top=1in, left=0.75in, right=0.75in, bottom=1in]{geometry}
\usepackage{amsfonts}
\journal{Journal of \LaTeX\ Templates}

\usepackage{listings}%

\begin{document}

\begin{frontmatter}

\title{Singular Lagrangians and the Faddeev-Jackiw
Formalism in Classical Mechanics}

\author{Jorge Mauricio Paulin Fuentes, Carlos Manuel López Arellano and Jaime Manuel Cabrera}
\address{ Divisi\'on Acad\'emica de Ciencias B\'asicas, Universidad Ju\'arez Aut\'onoma de Tabasco,\\ 
 Km 1 Carretera
Cunduac\'an-Jalpa, Apartado Postal 24, 86690 Cunduac\'an, Tabasco, M\'exico}


\cortext[mycorrespondingauthor]{jaime.manuel@ujat.mx, jorge.paulin@ujat.mx, carlosmazzepa@gmail.com}


\begin{abstract}
Classical mechanical systems with internal constraints will be examined using the extended symplectic formalism of Faddeev-Jackiw. We will derive the generalized brackets of the theory and the corresponding equations of motion. The investigation will include the analysis of a gauge system, leading to the identification of associated gauge transformations. The obtained results will be compared with those obtained through the Dirac-Bergmann algorithm, as previously documented in the literature. The symplectic approach turns out to be more simpler and efficient compared to the Dirac methods in its mathematical operations, and easy for computer implementation. 

It will be demonstrated that the symplectic approach proves to be simpler and more economical compared to the Dirac method.
\end{abstract}


\end{frontmatter}

\linenumbers

\section{Introduction}

Physical systems often exhibit constraints. In elementary classical mechanics \cite{Lan,gold}, the motion of these objects is typically restricted by externally applied constraints\footnote{Example: $\phi(q)=\phi(x,y)=x^{2}+y^{2}-l^{2}=0$.}, expressed as $\phi(q)=0$. Additionally, internal constraints, known as inherent constraints (derived from a singular Lagrangian) \cite{Teitelboim}, can also arise. It is important to note that the conventional quantization method is not directly applicable to dynamic systems with this type of constraint. 

Systems governed by singular Lagrangians\footnote{
This means that the matrix of second derivatives of $L(q,\dot{q})$ with respect to the velocities is non-invertible.}, known as singular systems, describe non-regular systems with constraints that indicate relationships among dynamic variables \cite{Teitelboim}. The methodology for addressing these systems was established by Paul Dirac \cite{Dirac} and Peter Bergmann \cite{Ber, Ber1}. They introduced a categorization of constraints within a system into primary, secondary, and tertiary constraints. The main goal is to classify these constraints as first and second class, with the latter being fundamental for constructing 'Dirac brackets,' essential elements in the quantization process \cite{Teitelboim, Sun, Hason,Routhe}. Conversely, gauge theories, characterized by first-class constraints in accordance with Dirac's constraint classification framework, play a crucial role in theoretical physics. They enable the exploration of fundamental interactions (electromagnetism, gravity, and nuclear forces) playing a crucial role in advancing our understanding of the underlying principles of the physical universe.




While this formalism fulfills its intended purpose, it can be tedious in certain cases, and the classification of constraints into first and second classes is not always straightforward
An alternative to the Dirac-Bergmann algorithm is the Faddeev-Jackiw method \cite{Faddeev}. In this formulation, all constraints are addressed without any additional categorization.
Besides, this geometric approach utilizes a symplectic structure and provides a direct derivation of generalized brackets from a matrix called the Faddeev-Jackiw symplectic matrix, applicable to regular systems. The extension of this method, dealing with constrained systems, was carried out by Barcelos-Neto and Wotzasek \cite{Bar,Bar1}. The symplectic approach simplifies the count of degrees of freedom, as all constraints have the same status, and relevant information is obtained through the invertible symplectic matrix, coinciding with brackets obtained in the Dirac method \cite{Liao,Huang}.


This work explores the dynamics of constrained systems using the Faddeev-Jackiw (FJ) formalism in classical mechanics, serving as a crucial preliminary step in comprehending the quantum version of such systems. This perspective proves valuable in various contexts, including field theory, gauge theories, and general relativity. While many works in the literature, particularly those focused on first-class constraints \cite{Huang1, Huang2, Bar3, Lee, Escalante, Ro, Jaime}, immediately delve into field theory, it's noteworthy that the FJ formalism is equally applicable to finite-dimensional systems. This is particularly advantageous as finite-dimensional systems are simpler compared to the complex field theories with an infinite number of degrees of freedom. Its application in mechanical systems is focused on toy models \cite{Anjali, Caro} to exemplify the functioning of the method.

Our analysis is specifically geared towards applying the Faddeev-Jackiw symplectic formalism to mechanical systems with a finite number of degrees of freedom, aiming to convey the fundamental concepts clearly and concisely.


In this investigation, we will utilize the Faddeev-Jackiw method \cite{Liao,Huang} to explore classical mechanical systems of the mass-spring type. These systems have been previously examined by David Brown \cite{Brown, Brown1} within the context of the Dirac-Bergmann formalism. The study aims to demonstrate the equivalence between the Faddeev-Jackiw and Dirac-Bergmann methods in analyzing these specific systems.

Section 2 presents various physical examples of singular systems built from familiar elements found in classical mechanics problems, such as point masses connected by massless springs, rods, strings, and pulleys. These aforementioned systems will be analyzed using the Faddeev-Jackiw formalism.
In Section 2.1, deals with the derivation
of constraints of a system of two springs coupled to a mass under the influence of gravity will be considered, using a modified version of the Faddeev-Jackiw formalism. Section 2.2 will focus on a pendulum with a movable coupling point through two springs. Both systems exhibit constraints, and generalized Faddeev-Jackiw brackets and equations of motion will be derived. Section 3 will analyze the Lagrangian proposed in \cite{Brown1}, a system with constraints and gauge symmetry, obtaining gauge transformation rules and generalized brackets. 

Our appendix provides the explicit computation of the generalized Faddeev-Jackiw brackets through a code implemented using the Wolfram Mathematica software.

\section{A Brief Review of Faddeev-Jackiw formalism}

\subsection{Compound spring}
To approach systems from the symplectic perspective of Faddeev-Jackiw, we start with a linearized Lagrangian of the form \cite{Faddeev}
\begin{equation}
\mathcal{L}^{(0)}=a^{(0)}_{i}\dot{\xi}^{(0)}_{i}-V^{(0)}, \quad i=1,\ldots,N.
\label{1}
\end{equation}

We begin with the Lagrangian of \cite{Brown1}, which consists of a mass $m$ fixed to a spring coupled with another spring, all under the influence of gravity with constants $k_1$ and $k_2$. The Lagrangian that describes the system is given by
\begin{equation}
     L=\frac{m}{2}(\dot{x}_{1}+\dot{x}_{2})^{2}+mg(x_{1}+x_{2})-\frac{k_{1}}{2}(x_{1}-l_{1})^{2}-\frac{k_{2}}{2}(x_{2}-l_{2})^{2}.
     \label{2}
\end{equation}
From the structure (\ref{1}), the Lagrangian (\ref{2}) is rewritten as follows
\begin{equation}
    \mathcal{L}^{(0)}=p_{1}(\dot{x}_{1}+\dot{x}_{2})-V^{(0)},
    \label{3}
\end{equation}
where 
\begin{equation}
    V^{(0)}=\dfrac{p_{1}^{2}}{2m}-mg(x_{1}+x_{2})+\dfrac{k_1}{2}(x_{1}-l_{1})^2+\dfrac{k_2}{2}(x_{2}-l_{2})^2,
    \label{4}
\end{equation}
is called symplectic potencial. The superscript $^{(0)}$ means initial Lagrangian. The equations of motion in the Faddeev-Jackiw approach are \cite{Faddeev}
\begin{equation}
f_{ij}^{(0)}\dot{\xi}^{j}=\dfrac{\partial V^{(0)}(\xi)}{\partial \xi^{i}},
    \label{5}
\end{equation}
where we define the symplectic matrix as 
\begin{equation}
    f^{(0)}_{ij}=\dfrac{\partial a^{(0)}_{j}}{\partial \xi_{i}}-\dfrac{\partial a^{(0)}_{i}}{\partial \xi_{j}}.
    \label{6}
\end{equation}
From (\ref{3}) we can identify the first set of symplectic variables $\xi^{(0)}_i=\{ x_{1},x_{2},p_{1}\}$, and the components of canonical 1-form are $a^{(0)}_i=\{ p_{1},p_{1},0 \}$. Therefore, the symplectic matrix has the following structure 
\begin{equation}
    f^{(0)}_{ij}=\begin{pmatrix}
     & \vline & x_1 & x_2 & p_1 \\
    \hline
    x_1 & \vline & 0 & 0 & -1 \\
    x_2 & \vline & 0 & 0 & -1 \\
    p_1 & \vline & 1 & 1 & 0
\end{pmatrix},
\label{7}
\end{equation}

where $f^{0}_{ij}$ is a singular matrix. Using the Barcelos-Netos approach \cite{Bar,Bar1} to treat systems with constraints. The matrix (\ref{7}) has a dimension of 3 and a rank of 2; therefore, the dimension of the null space is 1. Now, we calculate the zero modes of the matrix (\ref{7}) and is give by $(v^{(0)}_1)^{T}=(-1 \quad 1 \quad 0)$,  where multiplying by equation (\ref{5}) the structure of the constraints is obtained as 
\begin{align}
    \Omega^{(0)}_{1}&=(v^{(0)}_{1})_{i}^{T} \dfrac{\partial V^{(0)}}{\partial \xi_{i}^{(0)}}, \nonumber \\ 
     \Omega_{1}^{(0)} & =k_{1}(x_{1}-l_{1})-k_{2}(x_{2}-l_{2})=0,
     \label{8}
\end{align}
this constraint coincides with the secondary constraint of the Dirac-Bergmann formalism which is analyzed by Brown \cite{Brown}. To ensure equivalence between the FJ formalism and the Dirac-Bergmann algorithm, we make the constraints satisfy the consistency condition \cite{Liao,Huang,Huang1}
\begin{equation}
    \dot{\Omega}^{(0)}_\alpha=\dfrac{\partial \Omega^{(0)}_\alpha}{\partial \xi^{i}}\dot{\xi}^{i}=0,
    \label{9}
\end{equation}
which allows us to know more restrictions. Combining (\ref{5}) and (\ref{9}), we obtain 
\begin{equation}
    \tilde{f}^{(1)}_{kj}\dot{\xi}^{j}=Z_{k}(\xi),
    \label{10}
\end{equation}
where 
\begin{equation}
    Z_{k}(\xi)=\begin{pmatrix}
        \dfrac{\partial V^{(0)}(\xi)}{\partial \xi^{i}} \\
        0
    \end{pmatrix},
    \label{11}
\end{equation}
and 
\begin{equation}
    \tilde{f}^{(1)}=\begin{pmatrix}
         f_{ij}^{(0)} \\
        \dfrac{\partial \Omega^{(0)}}{\partial \xi^{i}} 
    \end{pmatrix}=
    \begin{pmatrix}
     & \vline & x_1 & x_2 & p_1 \\
    \hline
    x_1 & \vline & 0 & 0 & -1 \\
    x_2 & \vline & 0 & 0 & -1 \\
    p_1 & \vline & 1 & 1 & 0 \\
    \dfrac{\partial \Omega^{(0)}}{\partial \xi^i}  & \vline & 1 & 1 & 0 
\end{pmatrix}.
\label{12}
\end{equation}
Even without being a square matrix, it has a zero mode, given by $(\tilde{v}^{(1)}_1)^{T}=(-1, 1, 0, 0)$, this is used in order to obtain further constraints, multiplying this mode by equation (\ref{10}) and evaluating in the restriction of the equation, gives 
\begin{equation}
     \Omega^{(1)}_1=(v^{(1)}_1)_k^{T} Z_{k}\mid_{\Omega=0},
     \label{13}
\end{equation}
obtaining an identity, therefore, all the restrictions have been found.

We can write the new Lagrangian as 
\begin{equation}
    \mathcal{L}^{(1)}=p_{1}(\dot{x}_{1}+\dot{x}_{2})+\dot{\lambda} (k_{1}(x_{1}-l_{1})-k_{2}(x_{2}-l_{2}))-V^{(1)},
    \label{14}
\end{equation}
where $V^{(1)}=V^{(0)}\mid_{\Omega^{(0)}}={p_{1}^{2}}/{2m}-mg(x_{1}+x_{2})+{1}/{2}(x_{1}-l_{1})^{2} (k_{1}+{k_{1}^{2}}/{k_{2}})$. The set of symplectic variables in its first iteration is $\xi^{(1)}_i=\{ x_{1},x_{2},p_{1},\lambda \}$ noting that $\lambda$ is considered as a new dynamic variable and the new coefficients of
1-forms are $a^{(1)}_i=\{ p_{1},p_{1},0,k_{1}(x_{1}-l_{1})-k_{2}(x_{2}-l_{2})  \}$, identifying these new elements, the symplectic matrix is
\begin{equation}
    {f}^{(1)}=
    \begin{pmatrix}
     & \vline & x_1 & x_2 & p_1 & \lambda \\
    \hline
    x_1 & \vline & 0 & 0 & -1 & k_1 \\
    x_2 & \vline & 0 & 0 & -1 & -k_2\\
    p_1 & \vline & 1 & 1 & 0 & 0 \\
    \lambda  & \vline & -k_1 & k_2 & 0 & 0  
\end{pmatrix}.
\label{15}
\end{equation}
Which is a non-singular matrix and calculating its inverse given as 
\begin{equation}
    [{f}^{(1)}]^{-1}=
    \begin{pmatrix}
     & \vline & x_1 & x_2 & p_1 & \lambda \\
    \hline
    x_1 & \vline & 0 & 0 & \dfrac{k_2}{k_{1}+k_{2}} & -\dfrac{1}{k_{1}+k_{2}} \\
    x_2 & \vline & 0 & 0 & \dfrac{k_1}{k_{1}+k_{2}} & \dfrac{1}{k_{1}+k_{2}}\\
    p_1 & \vline & -\dfrac{k_2}{k_{1}+k_{2}} & \dfrac{k_1}{k_{1}+k_{2}} & 0 & 0\\
    \lambda  & \vline & \dfrac{1}{k_{1}+k_{2}} & -\dfrac{1}{k_{1}+k_{2}} & 0 &0  
\end{pmatrix}.
\label{16}
\end{equation}
The elements of the inverse matrix are identified with the generalized Faddeev brackets as follows 
\begin{equation}
    \{ \xi^{(1)}_i, \xi^{(1)}_j  \}_{FJ}=[{f}_{ij}^{(1)}]^{-1},
    \label{17}
\end{equation}
 therefore, from (\ref{16}) we obtain the following relations
 \begin{align}
     \{ x_{1},p_{1}  \}_{FJ}&=\dfrac{k_2}{k_{1}+k_{2}}, \nonumber \\
     \{ x_{2},p_{1}  \}_{FJ}&=\dfrac{k_1}{k_{1}+k_{2}}, \nonumber \\
     \{ x_{1},\lambda  \}_{FJ}&=-\dfrac{1}{k_{1}+k_{2}}, \nonumber \\ 
      \{ x_{2},\lambda  \}_{FJ}&=\dfrac{1}{k_{1}+k_{2}}. 
 \end{align}
From (\ref{5}) the equations of motion are calculated as
\begin{align}
    \dot{x}_{1}&=\{ x_{1},p_{1} \}_{FJ}\dfrac{\partial V^{(1)}}{\partial p_{1}}=\dfrac{k_2}{k_{1}+k_{2}} \dfrac{p_{1}}{m}, \nonumber \\
    \dot{p}_{1}&=\{ p_{1},x_{1} \}_{FJ}\dfrac{\partial V^{(1)}}{\partial x_{1}}+\{ p_{1},x_{2} \}_{FJ}\dfrac{\partial V^{(1)}}{\partial x_{2}}=mg-k_{1}(x_{1}-l_{1}), \nonumber \\
    \dot{x}_{2}&=\{ x_{2},p_{1} \}_{FJ}\dfrac{\partial V^{(1)}}{\partial p_{1}}=\dfrac{k_1}{k_{1}+k_{2}} \dfrac{p_{1}}{m},
    \label{19}
\end{align}
where the equations related to $\lambda$ have been omitted since, as mentioned in \cite{Huang1,Huang2,Anjali} the auxiliary variables that allow us to deform the symplectic matrix so that it can be invertible, are not real physical fields and are only combinations of the actual physical fields. Of the constrained (\ref{8}), it can be solved for some variables which will only allow two equations of (\ref{19}) to be solved, which in this case, if it is solved for $x_2$, the solutions are as follows 
\begin{align}
    x_{1}(t)&=C_{2} \cos{\omega t}+\dfrac{\omega}{k_1} C_{1} \sin{\omega t}+l_{1}+\dfrac{mg}{k_1}, \nonumber\\
    p_{1}(t)&=C_{1} \cos{\omega t}-\dfrac{k_1}{\omega}C_{2} \sin{\omega t}, \label{20}
\end{align}
with $\omega=\sqrt{k_{1}k_{2}/m(k_{1}+k_{2})}$. We worked with three dynamic variables $(x_1,x_2,p_1)$, and in the analysis a restriction was obtained $\Omega^{(0)}_1$, therefore, the count of degrees of freedom for this problem results in two physical degrees of freedom. 

\subsection{Pendulum and two springs}
The next problem is about a pendulum which is connected to two springs, the mechanical arrangement can be seen in \cite{Brown} since it was analyzed by Brown using the Dirac-Bergmann algorithm. The lagrangian that describes the system is given as
\begin{equation}
    L=\dfrac{m}{2}(\dot{x}^{2}+\dot{y}^2+l^{2}\dot{\theta}^{2})+ml (\dot{x}\cos{\theta}+\dot{y}\sin{\theta})\dot{\theta}-mg(y-l \cos{\theta})-k(x^{2}+y^{2}+d^{2}),
    \label{21}
\end{equation}
linearizing using the canonical moments as auxiliary variables
\begin{equation}    \mathcal{L}^{(0)}=p_{x}\dot{x}+p_{y}\dot{y}+(p_{x} l \cos{\theta}+p_{y} l \sin{\theta})\dot{\theta}-V^{(0)}, 
\label{22}
\end{equation}
where
\begin{equation}
    V^{(0)}=\dfrac{1}{2m}\left( p_{x}^{2}+p_{y}^{2} \right)+k(x^{2}+y^{2}+d^{2})+mg(y-l \cos{\theta}).
    \label{23}
\end{equation}
now we can identify from (\ref{22}) the symplectic variables as $\xi^{(0)}=\{ x,y,\theta,p_{x},p_{y} \}$ and the coefficients $a^{(0)}=\{p_{x},p_{y},p_{x} \cos{\theta}+p_{y} l \sin{\theta},0,0  \}$. Therefore, the simplex matrix is given by
\begin{equation}
    {f}^{(0)}=
    \begin{pmatrix}
     & \vline & x & y & \theta & p_{x} & p_{y} \\
    \hline
    x & \vline & 0 & 0 & 0 &-1 & 0\\
    y & \vline & 0 & 0 & 0 & 0 & -1\\
    \theta & \vline & 0 & 0 & 0 & -l \cos{\theta} & -l \sin{\theta}\\
    p_{x}  & \vline & 1 & 0 & l \cos{\theta} & 0 & 0 \\
    p_{y} & \vline & 0 & 1 & l \sin{\theta} & 0 & 0
\end{pmatrix},
\label{24}
\end{equation}
in which, it is a singular matrix. In this case obtaining the associated zero modes, which are $(v^{(0)}_{1})^{T}=(-l \cos{\theta},-l \sin{\theta},1, 0,0)$, and the constrained will have the following structure 
\begin{align}
    \Omega^{(0)}_{1}&=(v^{(0)}_{1})_i^{T} \dfrac{\partial V^{(0)}}{\partial \xi_{i}^{(0)}}=0\nonumber \\
    \Omega^{(0)}_{1}&=2kl (x \cos{\theta}+y \sin{\theta})=0.
    \label{25}
\end{align}
The consistency condition on the restrictions leads to calculating the non-square matrix $\tilde{f}^{(1)}_{kj}$ that has the structure as
\begin{equation}
    \tilde{f}^{(1)}=
    \begin{pmatrix}
     & \vline & x & y & \theta & p_{x} & p_{y} \\
    \hline
    x & \vline & 0 & 0 & 0 &-1 & 0\\
    y & \vline & 0 & 0 & 0 & 0 & -1\\
    \theta & \vline & 0 & 0 & 0 & -l \cos{\theta} & -l \sin{\theta}\\
    p_{x}  & \vline & 1 & 0 & l \cos{\theta} & 0 & 0 \\
    p_{y} & \vline & 0 & 1 & l \sin{\theta} & 0 & 0 \\ 
    \dfrac{\partial \Omega^{(0)}}{\partial \xi} & \vline & 2kl \cos{\theta} & 2kl \sin{\theta} & 2kl(y \cos{\theta}-x \sin{\theta}) & 0 & 0 
\end{pmatrix},
\label{26}
\end{equation}
which has the zero mode given by $(\tilde{v}^{(1)}_1)^{T}=(-l \cos{\theta},-l \sin{\theta},1, 0,0,0)$, If the multiplication is carried out with the potential $Z_k$, this generates an equality, therefore no new restrictions are found.

The new Lagrangian has the following form
\begin{equation}     \mathcal{L}^{(1)}=p_{x}\dot{x}+p_{y}\dot{y}+(p_{x} l \cos{\theta}+p_{y} l \sin{\theta})\dot{\theta}+\dot{\lambda}2kl(x \cos{\theta}+y \sin{\theta})-V^{(1)},   \label{27}
\end{equation}
where 
\begin{equation}
    V^{(1)}=V^{(0)}\mid_{\Omega^{(0)}}=\dfrac{1}{2m}\left( p_{x}^{2}+p_{y}^{2} \right)+k(x^{2}+y^{2}+d^{2})+mgy(1+\dfrac{l}{\sqrt{x^2+y^2}}),
    \label{28}
\end{equation}
and identifying the new set of symplectic variables as $\xi^{(1)}=\{ x,y,\theta,p_{x},p_{y},\lambda \}$ and the new  coefficients of 1-form $a^{(1)}=\{p_{x},p_{y},p_{x} \cos{\theta}+p_{y} l \sin{\theta},0,0,2kl(x \cos{\theta}+y \sin{\theta})  \}$. With this new information, we can construct the following symplectic matrix
\begin{equation}
    {f}^{(1)}=
    {
    \begin{pmatrix}
     & \vline & x & y & \theta & p_{x} & p_{y} & \lambda \\
    \hline
    x & \vline & 0 & 0 & 0 &-1 & 0 & 2kl \cos{\theta}\\
    y & \vline & 0 & 0 & 0 & 0 & -1 & 2kl \sin{\theta}\\
    \theta & \vline & 0 & 0 & 0 & -l \cos{\theta} & -l \sin{\theta} & -2kl(y \cos{\theta}-x \sin{\theta})\\
    p_{x}  & \vline & 1 & 0 & l \cos{\theta} & 0 & 0 & 0\\
    p_{y} & \vline & 0 & 1 & l \sin{\theta} & 0 & 0 & 0 \\ 
    \lambda & \vline & -2kl \cos{\theta} & -2kl \sin{\theta} & -2kl(y \cos{\theta}-x \sin{\theta}) & 0 & 0 & 0
\end{pmatrix},
}
\label{29}
\end{equation}
it is a non-singular matrix, so, calculating its inverse as
\begin{equation}
    [f^{(1)}]^{-1}=
    \begin{pmatrix}
        \textbf{0} & \mathcal{F} \\
        -[\mathcal{F}^{T}] & \mathbf{0}
    \end{pmatrix},
    \label{30}
\end{equation}
where 
\begin{equation}
    \mathcal{F} = \begin{pmatrix}
        \frac{\sin{\theta}  (l \sin {\theta} +x)-y \cos {\theta}}{l+x \sin {\theta }-y \cos \theta } &  -\frac{l \sin \theta  \cos \theta }{l+x \sin \theta -y \cos \theta } & \frac{1}{-2 k l \sec \theta -2 k x \tan \theta +2 k y} \\ 
        -\frac{l \sin \theta  \cos \theta }{l+x \sin \theta -y \cos \theta } & 1-\frac{l \sin ^2{\theta }}{l+x \sin \theta -y \cos \theta }  &  \frac{1}{-2 k l \csc \theta -2 k x+2 k y \cot \theta } \\
        \frac{1}{l \sec \theta +x \tan \theta -y} & \frac{1}{l \csc \theta +x-y \cot \theta } & \frac{1}{2 k l^2+2 k l x \sin \theta -2 k l y \cos \theta } 
    \end{pmatrix}.
    \label{31}
\end{equation}
If we take $\sin{\theta}={x}/{\sqrt{x^{2}+y^{2}}}, \cos{\theta}=-{y}/{\sqrt{x^{2}+y^{2}}}, r=\sqrt{x^{2}+y^{2}}$, the relationship between the elements of the inverse matrix and the generalized Faddeev-Jackiw brackets simplify as
\begin{align}
     \{x,p_{x}  \}_{FJ}&= \dfrac{r+lx^{2}/r^{2}}{r+l}, & \{x,p_{y}  \}_{FJ}&=\dfrac{lxy/r^{2}}{r+l}, \nonumber \\ 
    \{y,p_{x}  \}_{FJ}&=\dfrac{lxy/r^{2}}{r+l},  & \{y,p_{y}  \}_{FJ}&=\dfrac{r+ly^{2}/r^{2}}{r+l}, \nonumber \\ \{\theta,p_{x}  \}_{FJ}&= -\dfrac{y}{r(r+l)}, & \{\theta,p_{y}  \}_{FJ}&= \dfrac{x}{r(r+l)},\nonumber \\
    \{x, \lambda \}_{FJ}&=\dfrac{y}{2kr(r+l)}, & \{y, \lambda \}_{FJ}&=-\dfrac{x}{2kr(r+l)},\nonumber \\
     \{\theta, \lambda \}_{FJ}&=\dfrac{1}{2kl(l+r)},
     \label{FJ1}
\end{align}
and if the restriction for $\theta$ is solved, the equations of motion are written as follows
\begin{align}
    \dot{x}&=\{x,p_{x}\}_{FJ}\dfrac{\partial V^{(1)}}{\partial p_{x}}+\{x,p_{y}\}_{FJ}\dfrac{\partial V^{(1)}}{\partial p_{y}}=\dfrac{lx(y p_{y}+x p_{x})+p_{x}r^{3}}{m r^{2}(r+l)}, \nonumber \\
    \dot{p}_{x}&=\{p_{x},x\}_{FJ}\dfrac{\partial V^{(1)}}{\partial x}+\{p_{x},y\}_{FJ}\dfrac{\partial V^{(1)}}{\partial y}=-2kx, \nonumber \\
     \dot{y}&=\{y,p_{x}\}_{FJ}\dfrac{\partial V^{(1)}}{\partial p_{x}}+\{y,p_{y}\}_{FJ}\dfrac{\partial V^{(1)}}{\partial p_{y}}=\dfrac{ly(y p_{y}+x p_{x})+p_{y}r^{3}}{m r^{2}(r+l)}, \nonumber \\
     \dot{p}_{y}&=\{p_{y},x\}_{FJ}\dfrac{\partial V^{(1)}}{\partial x}+\{p_{y},y\}_{FJ}\dfrac{\partial V^{(1)}}{\partial y}=-2ky-mg,
\end{align}

\section{Gauge system}

Brown \cite{Brown1} proposes a model arising from the need to provide an example that encompasses all the logical steps of the Dirac-Bergmann formalism. This need arises because the majority of examples in the literature focus on one or two specific steps, complicating the task for students to integrate these fragments for a complete understanding. Brown advocates for a comprehensive example that not only makes it easier to understand each step individually but also allows for the smooth connection of all elements of the algorithm. This approach is particularly beneficial for those who learn best through complete examples, offering a holistic view of the process. In this section, our goal is to analyze the model \cite{Brown1} using the Faddeev-Jackiw formalism.

The system analyzed in this section is deﬁned by the Lagrangian \cite{Brown1}

\begin{equation}
    L=\dfrac{1}{2} \left[ (q_{1}+\dot{q}_{2}+\dot{q}_{3})^{2}+(\dot{q}_{4}-\dot{q}_{2})^{2}+(q_{1}+2q_{2})(q_{1}+2q_{4})  \right],
    \label{34}
\end{equation}
therefore, linearizing the Lagrangian using the canonical moments as auxiliary
\begin{equation}   
\mathcal{L}^{(0)}=(p_{3}-p_{4})\dot{q}_{2}+p_{3}\dot{q}_{3}+p_{4}\dot{q}_{4}-V^{(0)},
\label{35}
\end{equation}
where 
\begin{equation}
    V^{(0)}=\dfrac{1}{2} \left[ p_{3}^{2}+p_{4}^{2}-2p_{3}q_{1}-(q_{1}+2q_{2})(q_{1}+2q_{4}) \right]
    \label{36}. 
\end{equation}
It can be identified from (\ref{35}) the simplectic variables given as $ \xi^{(0)}=\{ q_{1},q_{2},q_{3},q_{4},p_{3},p_{4} \}$ and we can identify the coefficients of the 1-canonical form $a^{(0)}=\{0, p_{3}-p_{4},p_{3},p_{4},0,0 \}$ such that the symplectic matrix is
\begin{equation}
       f^{(0)}=
\begin{pmatrix}
  & \vline & q_{1} & q_{2} & q_{3} & q_{4} & p_{3} & p_{4} \\ \hline
q_{1} & \vline & 0 & 0 & 0 & 0 & 0 & 0 \\
q_{2} & \vline & 0 & 0 & 0 & 0 & -1 & 1 \\
 q_{3} & \vline & 0 & 0 & 0 & 0 & -1 & 0 \\
 q_{4} & \vline & 0 & 0 & 0 & 0 & 0 & -1 \\
 p_{3} & \vline & 0 & 1 & 1 & 0 & 0 & 0 \\
 p_{4} & \vline & 0 & -1 & 0 & 1 & 0 & 0 \\
\end{pmatrix}
\label{37},
\end{equation}
where this matrix is singular. Calculating the null vectors which are $(v^{(0)}_1)^{T}=(0,1,-1,1,0,0)$ and 
$(v^{(0)}_2)^{T}=(1,0,0,0,0,0)$, therefore we obtain the following restrictions
\begin{align}
    \Omega_{1}^{(0)}&=2(q_{1}+q_{2}+q_{4}), \nonumber \\
    \Omega_{2}^{(0)}&=q_{1}+q_{2}+q_{4}+p_{3}, 
    \label{38}
\end{align}
this leads us to construct the non-square matrix $\tilde{f}_{kj}^{(1)}$, which has the structure
\begin{equation}
       \tilde{f}^{(1)}=
       {
\begin{pmatrix}
  & \vline & q_{1} & q_{2} & q_{3} & q_{4} & p_{3} & p_{4} \\ \hline
q_{1} & \vline & 0 & 0 & 0 & 0 & 0 & 0 \\
q_{2} & \vline & 0 & 0 & 0 & 0 & -1 & 1 \\
 q_{3} & \vline & 0 & 0 & 0 & 0 & -1 & 0 \\
 q_{4} & \vline & 0 & 0 & 0 & 0 & 0 & -1 \\
 p_{3} & \vline & 0 & 1 & 1 & 0 & 0 & 0 \\
 p_{4} & \vline & 0 & -1 & 0 & 1 & 0 & 0 \\
 \dfrac{\partial \Omega^{(0)}_1}{\partial \xi} & \vline &  2 & 2 & 0 & 2 & 0 & 0 \\ 
 \dfrac{\partial \Omega^{(0)}_2}{\partial \xi} & \vline &  1 & 1 & 0 & 1 & 1 & 0
\end{pmatrix}
}
\label{39},
\end{equation}
from this matrix we have the following zero modes that are $(\tilde{v}^{(1)}_1)^{T}=(0,0,2,0,0,0,-1,2),$ $(\tilde{v}^{(1)}_2)^{T}=(0,1,-1,1,0,0,0,0),$ and $  (\tilde{v}^{(1)}_3)^{T}=(1,0,0,0,0,0,0,0)$. These zero modes do not generate new constraints, therefore the extended Lagrangian is given as
\begin{equation}
    \mathcal{L}^{(1)}=(p_{3}-p_{4})\dot{q}_{2}+p_{3}\dot{q}_{3}+p_{4}\dot{q}_{4}+2 \dot{\lambda}^{1}(q_{1}+q_{2}+q_{4})+\dot{\lambda}^{2}(q_{1}+q_{2}+q_{4}+p_{3})-V^{(1)},
    \label{40}
\end{equation}
where 
\begin{equation}
    V^{(1)}=\dfrac{1}{2}[\ p_{4}^{2}-(q_{1}+2q_{2})(q_{1}+2q_{4})]\
    \label{41},
\end{equation}
and identifying the new set of symplectics variables as $ \xi^{(1)}=\{q_{1},q_{2},q_{3},q_{4},p_{3},p_{4},\lambda^1,\lambda^2 \}$, the coefficients of the 1-forms are $a^{(1)}=\{ p_{3}-p_{4},p_{3},p_{4},0,0,0,2(q_{1}+q_{2}+q_{4}),q_{1}+q_{2}+q_{4}+p_{3} \}$, then the iterated symplectic matrix is
\begin{equation}
       {f}^{(1)}=
       {
\begin{pmatrix}
  & \vline & q_{1} & q_{2} & q_{3} & q_{4} & p_{3} & p_{4} & \lambda^1 & \lambda^2 \\ \hline
q_{1} & \vline & 0 & 0 & 0 & 0 & 0 & 0 & 2 & 1\\
q_{2} & \vline & 0 & 0 & 0 & 0 & -1 & 1 & 2 & 1  \\
 q_{3} & \vline & 0 & 0 & 0 & 0 & -1 & 0 & 0 & 0\\
 q_{4} & \vline & 0 & 0 & 0 & 0 & 0 & -1 & 2 & 1\\
 p_{3} & \vline & 0 & 1 & 1 & 0 & 0 & 0 & 0 & 1\\
 p_{4} & \vline & 0 & -1 & 0 & 1 & 0 & 0 & 0 & 0\\
 \lambda^1 & \vline &  -2 & -2 & 0 & -2 & 0 & 0 & 0 &0 \\ 
 \lambda^2 & \vline &  -1 & -1 & 0 & -1 & 0 & 0 & 0 & 0
\end{pmatrix}
}
\label{42}.
\end{equation}
Despite finding all the internal restrictions of the system, the symplectic matrix remains singular. This indicates the presence of gauge symmetry, therefore, following the Montani-Wotzasek \cite{Montani} formalism for gauge system, the gauge must be fixed, we choose the gauge condition given by Brown in \cite{Brown1}
\begin{align}
    \Phi_{1}&=q_{1}-q_{2}=0, \nonumber \\
    \Phi_{2}&=q_{3}+p_{4}=0.
    \label{43}
\end{align}
Introducing this new information into the Lagrangian (\ref{40})
\begin{align}
    \mathcal{L}^{(2)}=(p_{3}-p_{4})\dot{q}_{2}+p_{3}\dot{q}_{3}+p_{4}\dot{q}_{4}+& 2 \dot{\lambda}^{1}(q_{1}+q_{2}+q_{4})+\dot{\lambda}^{2}(q_{1}+q_{2}+q_{4}+p_{3}) \nonumber \\ 
    &\quad +\dot{\eta}^{1}(q_{1}-q_{2})+\dot{\eta}^{2}(q_{3}+p_{4})-V^{(2)},
    \label{44}
\end{align}
where 
\begin{equation}
    V^{(2)}=\dfrac{1}{2}[\ p_{2}^{2}+\dfrac{9}{4}q_{4}^{2} ]\ ,
    \label{45}
\end{equation}
which again identifying the symplectic variables $\xi^{(1)}=\{q_{1},q_{2},q_{3},q_{4},p_{3},p_{4},\lambda^1,\lambda^2,\eta^1,\eta^2 \}$ and the new coefficients are $a^{(1)}=\{ p_{3}-p_{4},p_{3},p_{4},0,0,0,2(q_{1}+q_{2}+q_{4}),q_{1}+q_{2}+q_{4}+p_{3},q_{1}-q_{2},q_{3}+p_{4} \}$. Now, the symplectic matrix is 
\begin{eqnarray}
f^{(2)}=
\left( 
\begin{array}{l|cccccccccc}
       & q_{1} & q_{2} & q_{3} & q_{4} & p_{3} & p_{4} & \lambda^1 & \lambda^2 & \eta^{1} & \eta^{2} \\ \hline
 q_{1} & 0 & 0 & 0 & 0 & 0 & 0 & 2 & 1 & 1 & 0 \\
 q_{2} & 0 & 0 & 0 & 0 & -1 & 1 & 2 & 1 & -1 & 0 \\
 q_{3} & 0 & 0 & 0 & 0 & -1 & 0 & 0 & 0 & 0 & 1 \\
 q_{4} & 0 & 0 & 0 & 0 & 0 & -1 & 2 & 1 & 0 & 0 \\
 p_{3} & 0 & 1 & 1 & 0 & 0 & 0 & 0 & 1 & 0 & 0 \\
 p_{4} & 0 & -1 & 0 & 1 & 0 & 0 & 0 & 0 & 0 & 1 \\
 \lambda^{1} & -2 & -2 & 0 & -2 & 0 & 0 & 0 & 0 & 0 & 0 \\
 \lambda^{2} & -1 & -1 & 0 & -1 & -1 & 0 & 0 & 0 & 0 & 0 \\
 \eta^{1} & -1 & 1 & 0 & 0 & 0 & 0 & 0 & 0 & 0 & 0 \\
 \eta^{2} & 0 & 0 & -1 & 0 & 0 & -1 & 0 & 0 & 0 & 0  \\
  \end{array}
\right)
\label{46}
\end{eqnarray}
which has an inverse of the form
\begin{eqnarray}
 [f^{(2)}]^{-1}=
 {\small
\left( 
\begin{array}{l|cccccccccc}
       & q_{1} & q_{2} & q_{3} & q_{4} & p_{3} & p_{4} & \lambda^1 & \lambda^2 & \eta^{1} & \eta^{2} \\ \hline
 q_{1} & 0 & 0 & \frac{1}{3} & 0 & 0 & -\frac{1}{3} & 0 & -\frac{1}{3} & -\frac{2}{3} & 0 \\
 q_{2} & 0 & 0 & \frac{1}{3} & 0 & 0 & -\frac{1}{3} & 0 & -\frac{1}{3} & \frac{1}{3} & 0 \\
 q_{3} & -\frac{1}{3} & -\frac{1}{3} & 0 & \frac{2}{3} & 0 & 0 & -\frac{1}{6} & \frac{1}{3} & 0 & -1 \\
 q_{4} & 0 & 0 & -\frac{2}{3} & 0 & 0 & \frac{2}{3} & -\frac{1}{2} & \frac{2}{3} & \frac{1}{3} & 0 \\
 p_{3} & 0 & 0 & 0 & 0 & 0 & 0 & \frac{1}{2} & -1 & 0 & 0 \\
 p_{4} & \frac{1}{3} & \frac{1}{3} & 0 & -\frac{2}{3} & 0 & 0 & \frac{1}{6} & -\frac{1}{3} & 0 & 0 \\
 \lambda^{1} & 0 & 0 & \frac{1}{6} & \frac{1}{2} & -\frac{1}{2} & -\frac{1}{6} & 0 & -\frac{1}{6} & \frac{1}{6} & -\frac{1}{2} \\
 \lambda^{2} & \frac{1}{3} & \frac{1}{3} & -\frac{1}{3} & -\frac{2}{3} & 1 & \frac{1}{3} & \frac{1}{6} & 0 & -\frac{1}{3} & 1 \\
 \eta^{1} & \frac{2}{3} & -\frac{1}{3} & 0 & -\frac{1}{3} & 0 & 0 & -\frac{1}{6} & \frac{1}{3} & 0 & 0 \\
 \eta^{2} & 0 & 0 & 1 & 0 & 0 & 0 & \frac{1}{2} & -1 & 0 & 0   \\
  \end{array}
\right)
}
\label{47}
\end{eqnarray}

writing the relationship of the elements of the inverse matrix with generalized Faddeev-Jackiw brackets, where the relationships between auxiliary variables are omitted
\begin{align}
    \{ q_{1},q_{3} \}_{FJ}&=\dfrac{1}{3}, & \{ q_{1},p_{4} \}_{FJ}&=-\dfrac{1}{3},  & \{ q_{2},q_{3} \}_{FJ}&=\dfrac{1}{3}, & \{ q_{2},p_{4} \}_{FJ}&=-\dfrac{1}{3}, \\
    \{ q_{3},q_{4} \}_{FJ}&=\dfrac{2}{3}, & \{ q_{4},p_{4} \}_{FJ}&=\dfrac{2}{3}. 
\end{align}
By solving the restrictions (\ref{38}) and (\ref{43}), you can obtain a description of the system based on two variables, which in this case is
\begin{align}
    \dot{q}_{4}&=\{ q_{4},p_{2} \}_{FJ} \dfrac{\partial V^{(2)}}{\partial p_{2}}=\{ q_{4},-p_{4} \}_{FJ} \dfrac{\partial V^{(2)}}{\partial p_{2}}=-\dfrac{2}{3}p_{2}, \nonumber \\
    \dot{p}_{2}&=\{ p_{2},q_{4} \}_{FJ}  \dfrac{\partial V^{(2)}}{\partial q_{4}}=\{ -p_{4},q_{4} \}_{FJ}  \dfrac{\partial V^{(2)}}{\partial q_{4}}=\dfrac{3}{2}q_{4},
\end{align}
thereby obtaining
\begin{align}
     q_{4}&=C_{1} \cos {t}-\frac{2}{3} C_{2} \sin {t}, \nonumber \\
    p_{2}&=C_2 \cos{t} + \dfrac{3}{2} C_{1} \sin{t}. \label{51}
\end{align}
From \cite{Montani} it is established that for gauge systems the zero modes associated with the non-invertible symplectic matrix are the generators of the gauge transformations, therefore the zero modes are
\begin{align}
    (v^{(1)}_1)^{T}&=(0,0,2,0,0,0,1,-2), \nonumber \\
    (v^{(1)}_2)^{T}&=(2,-1,1,-1,0,0,0,0).
    \label{52}
    \end{align}
The infinitesimal gauge transformations of the coordinates are given as \cite{Montani, Anjali}
\begin{equation}
    \delta \xi_{i}^{(1)}=\sum_{\alpha} v^{(1)}_{i \alpha}\epsilon^{\alpha},
    \label{53}
\end{equation}
where $\epsilon^{\alpha}$ is a set of infinitesimal gauge arbitrary time parameters. So, the gauge transformation forms for this system are
\begin{equation}
    \delta  q_{1}=2 \epsilon^{2}(t), \quad \delta q_{2}=-\epsilon^{2}(t), \quad \delta  q_{3}=2\epsilon^{1}(t)+\epsilon^{2}(t), \quad \delta  q_{4}=-\epsilon^{2}(t),
\end{equation}
these rules keep the Lagrangian and the equations of motion invariant under the set of gauge transformations.

\section{Conclusions and prospects}


In this paper, we analyze the physical constraints of mass-spring mechanical systems using the Faddeev–Jackiw formalism in Section 2. Our investigation reveals that incorporating all constraints into the first-order Lagrangian renders the two-form symplectic matrix nonsingular. Consequently, obtaining the Generalized Faddeev-Jackiw brackets from the inverse of this nonsingular matrix allows for the derivation of equations of motion. 
For the system discussed in Section 3, gauge transformation rules are obtained from the null vectors associated with the non-invertible symplectic matrix (\ref{42}). These rules keep both the Lagrangian and the equations of motion invariant. It is noteworthy that the results obtained coincide with those reported in \cite{Brown, Brown1}, establishing equivalence with the Dirac-Bergmann algorithm. This shows that the symplectic approach does not require extensive calculations or classifications of constraints; the only difficulty lies in inverting the symplectic matrix.

Therefore, it can be stated that the Faddeev-Jackiw method is a great alternative for handling singular systems, proving to be more effective. Once the bracket structures are obtained, they can be promoted to commutators for quantization \cite{Barbosa}
\begin{equation}
    \{ \xi_i, \xi_j  \}_{FJ}= -\frac{i}{\hbar}[\hat{\xi_i},\hat{\xi_j}].
\end{equation}
There are also other quantization approaches, such as through path integrals \cite{Toms,Liao1}. It is hoped that the quantum aspects of the systems can be developed, emphasizing the transition from classical to quantum systems in a more effective manner.

\section*{Acknowledgements}
The authors welcome the support of the Universidad Juárez Autónoma de Tabasco for providing a suitable work environment while this research was carried out. J.M.C. also thanks CONACYT for their support through a grant for postdoctoral studies under grant No. 3873825. We thank E. Chan-López for the discussions and insights regarding the work.

\section*{Data Availability Statement}

No Data associated in the manuscript.

\section*{Appendix: Faddeev-Jackiw Code in Wolfram Mathematica for Regular Systems}

In classical mechanics, the Faddeev-Jackiw approach offers a powerful framework for understanding constrained systems and symmetries. Originating from the study of gauge theories and constrained dynamics, the Faddeev-Jackiw formalism provides a systematic method for analyzing constrained systems, offering insights into their dynamics and symplectic structures. The construction of the Faddeev-Jackiw matrix, which encapsulates the symplectic structure of such systems, plays a fundamental role in this formalism. This matrix not only characterizes the constraints and their interaction, but also facilitates calculations of physical observables and symplectic transformations. The approach presented here can be found in the foundational works in this field (see \cite{Faddeev, Bar}).

Using a symbolic computing language to address the Faddeev-Jackiw method accelerates the analysis of constrained systems, particularly in cases where the resulting Faddeev-Jackiw matrix extends beyond four dimensions with intricate matrix elements. By adopting symbolic calculation techniques, errors are minimized in a streamlined analytical process, facilitating a deeper comprehension of the underlying dynamics. It's worth noting that the detailed implementation of the algorithm in a computational environment, specifically using Wolfram Mathematica. Below, we present pseudocode that describes the algorithmic steps involved in constructing the Faddeev-Jackiw matrix, offering a concise representation of the underlying mathematics.

\begin{algorithm*}
\caption{Construction of Faddeev-Jackiw Matrix}
\begin{algorithmic}[1]
    \Function{FaddeevJackiwOperator}{$i$, $j$, $a$, $\xi$}
        \If{$i = j$}
            \State \Return $0$
        \ElsIf{$i > j$}
            \State \Return $-\left(\displaystyle\frac{\partial a[i]}{\partial \xi[j]} - \frac{\partial a[j]}{\partial \xi[i]}\right)$
        \Else
            \State \Return $\displaystyle\frac{\partial a[j]}{\partial \xi[i]} - \frac{\partial a[i]}{\partial \xi[j]}$
        \EndIf
    \EndFunction
    
    \Function{FaddeevJackiwMatrix}{$f$, $coords$}
        \If{$\textproc{Length}(f) \neq \textproc{Length}(coords)$}
            \State \Return unevaluated
        \Else
            \If{$\textproc{det}(fjm) \neq 0$}
                \State \Return $fjm$, $\textproc{Inverse}(fjm)$
            \Else
                \State \Return $fjm$, $\{\}$
            \EndIf
        \EndIf
    \EndFunction
\end{algorithmic}
\end{algorithm*}

To check all results of the generalized brackets of Faddeev-Jackiw, the code structure written in Mathematica is outlined, followed by its application to a specific example from Section 2.2. First, the operator \texttt{FaddeevJackiwOperator} is defined to compute the matrix elements in the following lines:

\begin{figure}[H]
		\flushleft
		\includegraphics[width=0.8\columnwidth]{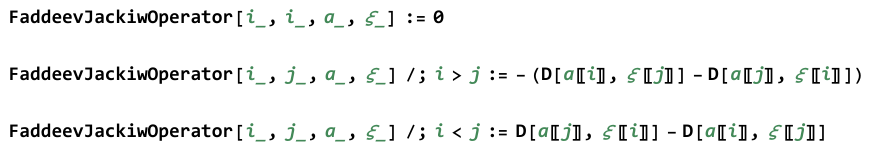}
		\label{codefaddeev}
	\end{figure}
 
The last step is to define the symplectic matrix \texttt{FaddeevJackiwMatrix} in the following way:

\begin{figure}[H]
		\flushleft
		\includegraphics[width=0.6\columnwidth]{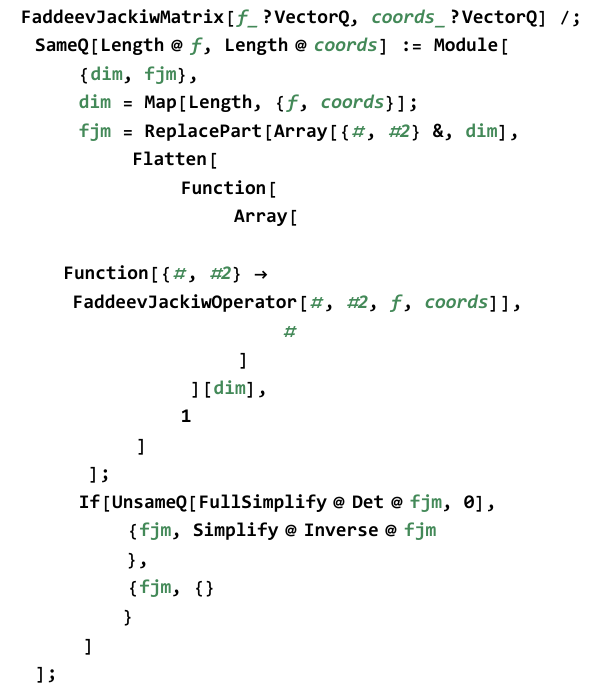}
		\label{codefaddeev1}
	\end{figure}

 To use the above top-level functions, proceed as follows:

\begin{itemize}
\item Set the one-form list \textbf{$a_i$} and the symplectic variables list \textbf{$\xi_i$}.
\item Execute \texttt{FaddeevJackiwOperator[$a_{i},\xi_{i}$]}.

\end{itemize}

We can identify from (\ref{22}) the symplectic variables as $\xi^{(0)}=\{x_{1},x_{2},x_{3},p_{1},p_{2}\}=\{ x,y,\theta,p_{x},p_{y} \}$ and the coefficients $a^{(0)}=\{p_{1},p_{2},p_{1}l\cos(x_{3})+p_{2}l\sin(x_{3}),0,0\}=\{p_{x},p_{y},p_{x}\cos{\theta}+p_{y} l\sin{\theta},0,0 \}$. In a Mathematica notebook session, we can assign the Mathematica input lists for the one-form and symplectic variables:

\begin{figure}[H]
		\flushleft
		\includegraphics[width=0.65\columnwidth]{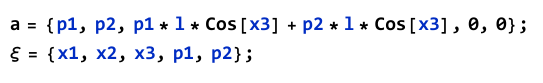}
		\label{codefaddeev2-1}
	\end{figure}

\vspace{-2.5em}
\begin{figure}[H]
		\flushleft
		\includegraphics[width=0.75\columnwidth]{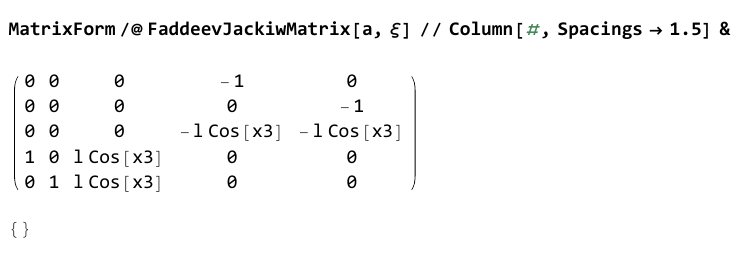}
		\label{codefaddeev2-2}
	\end{figure}

The previous output is a list composed by two sublists. The first list refers to the matrix elements of the symplectic matrix, as shown in reference (\ref{24}), while the second list displays the matrix inverse. If singularity occurs, as in our example, the algorithm determines that the symplectic matrix is not invertible and the displayed result corresponds to an empty list "$\{\}$".

After identifying all constraints within the Faddeev-Jackiw theoretical framework, as mentioned earlier, we proceeded to develop a new Lagrangian that considers them (\ref{27}). This newly formulated Lagrangian not only integrates these constraints effectively but also facilitates the creation of an extended symplectic matrix, using an updated set of symplectic variables and one-forms. The new one-forms and symplectic variables are given by $\xi^{(1)}=\{x_{1},x_{2},x_{3},p_{1},p_{2},x_4\}=\{ x,y,\theta,p_{x},p_{y},\lambda \}$ and the coefficients $a^{(1)}=\{p_{1},p_{2},p_{1}l\cos x_{3}+p_{2}l\sin x_{3},0,0,2kl(x\cos x_{3}+y\sin x_{3}\}=\{p_{x},p_{y},p_{x}\cos{\theta}+p_{y} l\sin{\theta},0,0,2kl(x\cos \theta+y\sin \theta) \}$. Finally, two lists have been reconstructed to incorporate the updated information from the extended Lagrangian:

\begin{figure}[htbp]
		\flushleft
		\includegraphics[width=1\columnwidth]{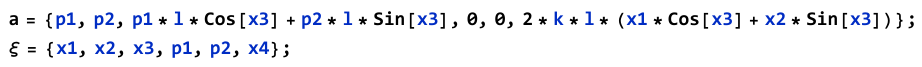}
		\label{codefaddeev3}
	\end{figure}

\vspace{-1.5em}
 After applying \texttt{FaddeevJackiwMatrix} to the above one-form and symplectic variable lists, we obtain the following ouputs:

\begin{figure}[H]
		\flushleft
		\includegraphics[width=0.9\columnwidth]{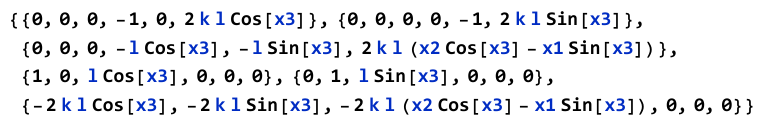}
		\label{codefaddeev4}
	\end{figure}

 \vspace{-3em} 

\begin{figure}[H]
		\flushleft
		\includegraphics[width=.9 \columnwidth]{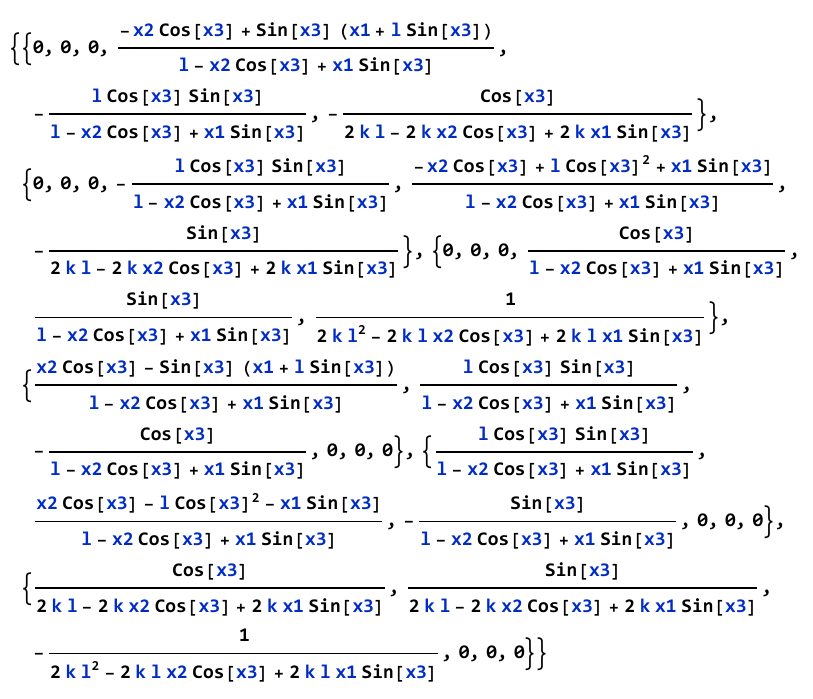}
		\label{codefaddeev5}
	\end{figure}

\vspace{-1em} 
The matrix elements obtained from the inverse using \texttt{FaddeevJackiwMatrix} operator are consistent with the results obtained from (\ref{FJ1}).

This perspective, rooted in both theoretical principles and computational methodologies, presents a powerful toolkit for exploring the dynamics of constrained systems, offering insights into their symplectic structures and facilitating further investigations in theoretical physics.


\end{document}